\documentclass[preprint]{aastex}

\newcommand{\ark}{Akn~564}	
\newcommand{\ngc}{NGC~3516}
\newcommand{\fxv}{\sigma_{XS}^2}
\newcommand{\et}{et al.\ }
\newcommand{\xte}{{\it RXTE}}
\newcommand{\rosat}{{\it ROSAT}}

\newcommand{\chandra}{{\it Chandra}}

\newcommand{\exosat}{{\it EXOSAT}}
\newcommand{\asca}{{\it ASCA}}
\newcommand{\hst}{{\it HST}}

\newcommand{\ls}
{\mathrel{\hbox{\rlap{\hbox{\lower4pt\hbox{$\sim$}}}\hbox{$<$}}}}
\newcommand{\gs}
{\mathrel{\hbox{\rlap{\hbox{\lower4pt\hbox{$\sim$}}}\hbox{$>$}}}}

\begin{document}

\title{X-ray Power Density Spectrum of the Narrow Line
Seyfert 1 Galaxy \ark}
\shorttitle{X-ray Variability of \ark}
\shortauthors{Pounds \et}
\author{ Ken Pounds\altaffilmark{1},
	Rick Edelson\altaffilmark{1,2},
	Alex Markowitz\altaffilmark{2},
	Simon Vaughan\altaffilmark{1,3} }

\email{kap@star.le.ac.uk}

\altaffiltext{1}{X-ray Astronomy Group; University of
Leicester; Leicester
LE1
7RH; United Kingdom}

\altaffiltext{2}{Astronomy Department; University of California; Los
Angeles, CA 90095-1562; USA}

\altaffiltext{3}{Institute of Astronomy; University of Cambridge; Cambridge CB3
0HA;
United Kingdom}

\begin{abstract}

Beginning in 1999 January, the bright, strongly variable Narrow-Line
Seyfert 1 (NLS1) galaxy \ark\ has been observed by \xte\ once every
$\sim$4.3~days.
It was also monitored every $\sim$3.2~hr throughout
2000 July.
These evenly-sampled observations have allowed the first quantitative
comparison
of long and short time-scale
X-ray variability in an NLS1 and the derivation of an X-ray Power Density
Spectrum (PDS). The variability amplitude in the short time-scale
light curve is very similar
to that in the long time-scale light curve, in marked
contrast to the stronger variability on longer time-scales which is
characteristic of ``normal" broad-line Seyfert 1s (BLS1s).
Furthermore, the \ark\ PDS power law cuts off at a frequency of
8.7~$\times$~10$^{-7}$~Hz corresponding to a timescale of $\sim$13~d,
significantly
{\it shorter}
than that seen in the PDS of \ngc, a BLS1 of comparable luminosity.

This result is consistent with NLS1s showing faster (as opposed to
larger amplitude) variations than BLS1s, 
providing further evidence that NLS1s harbour lower mass black
holes than BLS1s of similar luminosity, accreting at a correspondingly
higher relative rate.

\end{abstract}


\section{ Introduction }
\label{intro}

A relatively recent development in the taxonomy of active galactic
nuclei (AGN) has been the emergence
of Narrow-Line
Seyfert 1 (NLS1) galaxies as an important sub-class.
NLS1s were originally identified on the basis of their optical
properties: exhibiting narrower permitted lines (H$\beta$ FWHM $ < 2000 $~km/s)
than
"normal" Broad-Line Seyfert 1s (BLS1s) and weaker
[O III]/H$\beta$ $< 3 $ than Seyfert 2s (Osterbrock \& Pogge
1985; Goodrich 1989) .
\rosat\ data showed these properties to be strongly
correlated with X-ray spectral slope, in the sense that NLS1s tend to have
steep soft X-ray spectra (Boller \et 1996).
In addition, many NLS1s are strongly variable on time-scales of hours or
less, with some NLS1s showing giant (factor of $\sim$100) X-ray flares
over time-scales of
days (e.g., Brandt \et 1999).
On these short time-scales the variability levels in NLS1s are typically
a factor of
$\sim$3--10
times larger than those seen in BLS1s of similar luminosity (Turner \et
1999; Leighly 1999).

A number of models have been proposed to explain the distinctive properties
observed in NLS1s.
The most widely accepted explanation is that NLS1s are accreting
near the Eddington
limit (Pounds \et 1995, Laor \et 1997), while BLS1s accrete at a 
lower rate. On this interpretation the
high accretion rate is directly responsible for the strong soft
X-ray emission, by enhanced thermal radiation from the accretion disc, while
the increased soft photon flux cools the hard X-ray source leading to the
steeper power law frequently seen at higher X-ray energies in NLS1s
(Brandt \et 1997).
It is the prospect that comparative study of NLS1s and BLS1s
will shed light on the accretion rate, one of the fundamental
parameters of an AGN, that makes the emergence of the new sub-class
of AGN of such interest and importance.

X-ray variability has been recognised as a powerful probe of the
central regions of
AGN since the
\exosat\ ``long-looks" first showed (McHardy 1988) that
rapid variability was
common in Seyfert galaxies, lending early support to the now-standard black
hole/accretion disc paradigm.
The \exosat\ data were found to be well-described by a
fluctuation power density spectrum (PDS) rising smoothly to lower
frequencies as a power law, $f^{-\alpha}$ where $\alpha=1-2$,
to a limit imposed by the maximum duration of the
$\sim$2-day observations (e.g., Lawrence \& Papadakis 1993).
Attempts to constrain a flattening or "cut-off" in the PDS
(as required to avoid the
variability power diverging) were made by combining \exosat\ data
with earlier X-ray satellite observations of NGC 5506 (McHardy 1988)
and NGC 4151 (Papadakis and McHardy 1995). These analyses yielded
evidence for a cut-off time-scale of several weeks; however, the
uneven sampling of these data made their reliability uncertain.
The observational situation was significantly improved with the launch of
\xte.  Taking advantage of the unique properties of this satellite,
Edelson and Nandra (1999) obtained evenly-sampled fluxes
of the
BLS1 \ngc\, on
long, medium and short time-scales, and combined these data to produce
a PDS
covering 4 decades in temporal frequency, finding a cut-off time-scale 
of $\sim$1~month.

Until now the only quantitative assessment of the rapid
X-ray variability of NLS1s has been based on measurements of the excess
variance, obtained from \asca\ observations, typically of 1 day duration.
Based on these
data, Turner \et (1999) and Leighly (1999) found the short-term
variability of NLS1s to be substantially greater than for BLS1s,
while in both cases the excess variance was anticorrelated with the
X-ray luminosity.
The present paper reports the results of the first extensive and
evenly-sampled X-ray
monitoring of an
NLS1, using \xte\ to monitor \ark\ over a
near 2-year period.
\ark\ is particularly well suited for this study, being the
brightest known NLS1
in the hard X-ray
sky ($ F({\rm 2-10~keV}) \approx 2-5 \times 10^{-11} $~erg cm$^{-2}$~sec
$^{-1}$; Vaughan \et 1999a), strongly variable (e.g., 50\% variations on
timescales of hours; Leighly 1999) and located well out of the
ecliptic plane (allowing monitoring by \xte\ throughout the year).
Subsequent to the start of our \xte\ monitoring campaign, \ark\ was
chosen for simultaneous monitoring with \hst, \asca, \chandra\
and the AGN Watch network of ground based optical telescopes.

\section{ Observations and Data Reduction }

Our X-ray monitoring was planned to obtain variability information 
spanning time-scales from
hours to months.
\ark\ was observed once every 4.3~days (= 64 orbits) from 1999 January 1
-- 2000 September 19 (the limit of the present analysis, although this
low-frequency monitoring is continuing), and once every $\sim$3.2~hr (= 2
orbits) from 2000 June 30 -- August 1.
(Twice-daily observations were also obtained during 1999 October 25 --
November 11, in support of the AGN Watch campaign; Schemmer \et in
preparation).
Our sampling parameters are summarized in Table~1.

\placetable{tab1}

The \xte\ Proportional Counter Array (PCA) consists of five collimated
Proportional Counter Units (PCUs), sensitive in a nominal
2--60~keV bandpass (Jahoda \et
1996).
At the start of the present campaign three of the PCUs (0, 1 and 2) were in
routine use.
After the gain settings of all five PCUs on board \xte\ were changed on
1999 March 22 this number was reduced to two (0 and 1), and then, from
2000 May 12, to one (1).
Data collected during each of the two gain epochs were extracted
separately using background model files appropriate to that gain epoch.
The present analysis is restricted to the 2--10~keV band, where the PCA is
most sensitive and the systematic errors are best understood. Data from
the top (most sensitive) layer of the PCU array were extracted using the
{\tt REX} reduction script\footnote{See {\tt
http://heasarc.gsfc.nasa.gov/docs/xte/recipes/rex.html}}, and background 
counts were estimated using the ``L7--240'' model\footnote{See {\tt
http://lheawww.gsfc.nasa.gov/$\sim$keith/dasmith/rossi2000/index.html}}.
Background-subtracted count rates were then obtained in each epoch to assign
a mean flux and error.
Count rates derived from data taken during the earlier gain setting
were scaled to account for the effect of the gain change by a factor
derived from observations of Cassiopeia A, and
then normalised to units of ct s$^{-1}$ PCU$^{-1}$. Figure 1 (top panel)
reproduces the full light curve.

\section{ Temporal Analysis }

Before performing the time series analyses described below, the
observed light
curve was resampled to provide two nearly-independent light curves,
one sampling short time-scales ($\sim$days) and the other, long time-scales
($\sim$months).
In Figure~1, the centre panel contains the full 20 months of data, sampled on
a grid as close to 4.3~d as possible.
Interpolating over the 9 data points
that were missing during this period and mapping to an even grid
yielded a total of 148 evenly-spaced flux points.
This had the effect of filtering out (not smoothing) most of the data in
the intensive periods around MJD 51480 and 51700.
The short time-scale light curve, shown in the bottom panel of Figure~1,
was obtained during the 32~d
intensive sampling period 2000 June 30 -- August 1.
Interpolating over the 13 data points missing during this period
yielded a total of 235 evenly-spaced flux points.

\begin{figure}
\vspace{15 cm}
\includegraphics{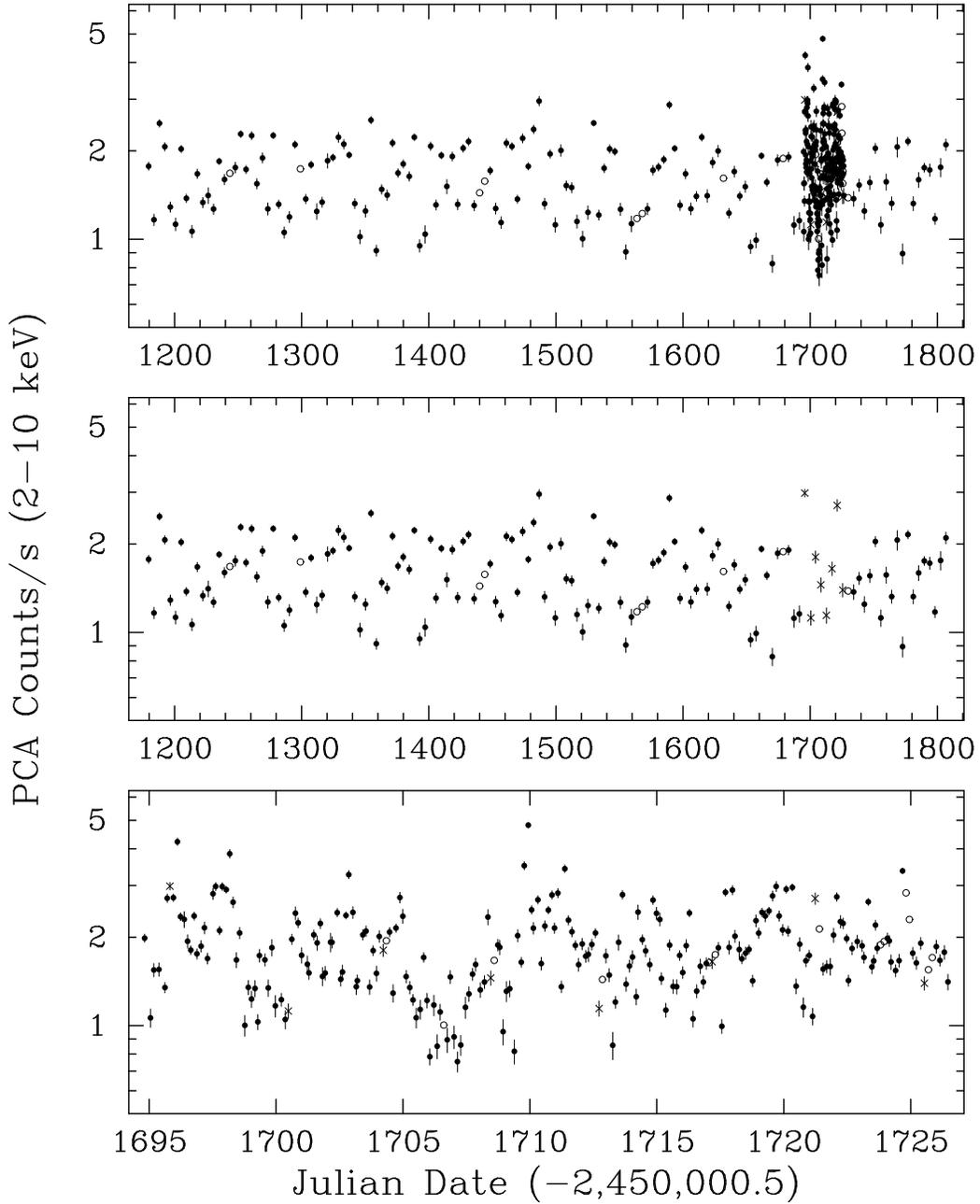}
\caption{Full band (2-10~keV) light curve (top), resampled on long
(4.3~d--20~months; centre ) and short (3.2~hr--31~d; bottom)
timescales.
These middle and lower light curves were produced by filtering and
interpolating the data in the top panel as described in the text.
Interpolated points are shown as open circles without error bars, and the 8
points common to both light curves as $\times$'s.}
\label{fig1}
\end{figure}

\placefigure{fig1}

\subsection{ Comparison of Long and Short Timescale Light Curves }

The long and short time-scale data are quantitatively compared in Table~2, 
where column~2
gives the mean count rates, and
column~4 the fractional excess variance, $\fxv$, defined as
\begin{equation}
\fxv = { S^2 - \langle \sigma_{err}^2 \rangle \over \langle X \rangle^2 },
\end{equation}
where $ \langle X \rangle $ is the mean flux, $ \langle \sigma_{err}^2
\rangle $ is the mean square error, and $S^2$ is the measured variance of
the light curve (e.g. Nandra \et 1997).
We have estimated ($\pm1 \sigma$) errors on the excess variance as $\pm
\fxv \sqrt{2/N}$, where N is the number of flux points, noting that for 
Gaussian fluctuations $S^2$ will
follow a $\chi^2$ distribution (Warwick, private communication).

\placetable{tab2}

The short time-scale light curve of \ark\ shows a marginally 
{\it larger} excess variance than the long
time-scale light curve, a
result that is strikingly different from that of BLS1s.
In a survey of \xte\ and \asca\ data for 8 BLS1s, 
Markowitz \&
Edelson (2001) found larger values of
$\fxv$ on long
($\sim$months) than on short ($\sim$hours) time-scales in every case.
This comparison strongly suggests that large amplitude X-ray variability is
more rapid, i.e., extends to significantly shorter time-scales, in \ark\
than in BLS1s
of comparable luminosity.

\subsection{Fluctuation Power Density Spectrum}

In order to better quantify the short-term variability of \ark\
the PDS was then constructed
following the prescription of Edelson \& Nandra (1999),
first determining separate short and long time-scale PDS and then
combining them to produce a single PDS.
The individual PDS were derived using a direct Discrete Fourier Transform
(Oppenheim \& Shafer 1975, Brillinger 1981).
The zero-power and next two (very noisy) lowest-frequency points of each
PDS were ignored and the remaining points binned every factor of 1.7
(0.23 in the logarithm).
Power-law models were then fitted separately to the long and short time-scale
PDS.
The slope and uncertainty, measured from an unweighted, least-squares fit
to the logarithmically binned data, and derived using the Welch window
function, are given in column~5 of Table~2.
These show a highly significant (10$\sigma$) systematic flattening from
short time-scales (with $ a = -0.96 \pm 0.07 $) to long time-scales ($ a =
-0.24 \pm 0.08 $).

The long and short time-scale PDS were then combined.
Unlike Edelson \& Nandra (1999), who allowed the relative normalization
between PDS to float as a free parameter, the absolute normalizations
were retained.
The combined PDS is shown in Figure~2.
A simple power law gave an unacceptable fit ($>$99.99\%), with a reduced
chi-squared of
$\chi^{2}_{\nu}= 3.06 $ for 13 degrees of freedom.

\begin{figure}
\vspace{12 cm}
\includegraphics{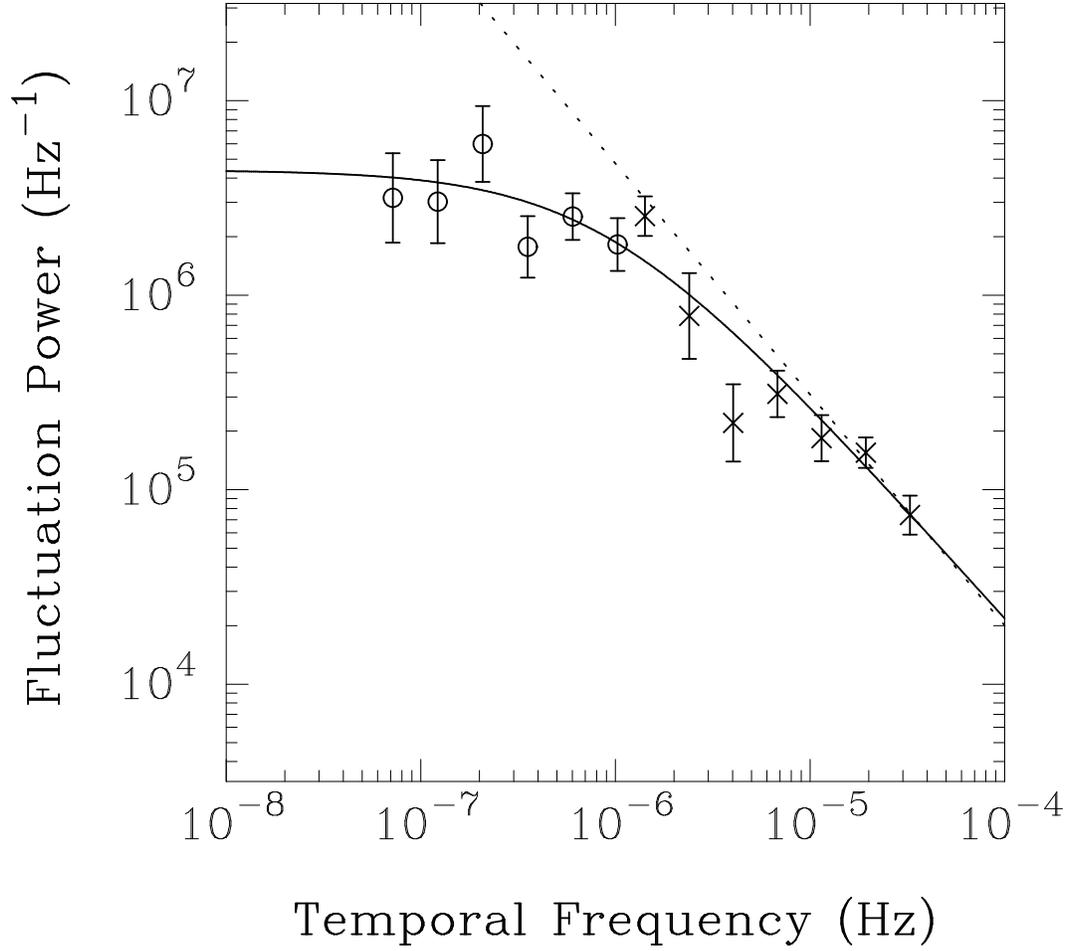}
\caption{The combined PDS of \ark.
The crosses refer to the PDS derived from the short term light curve
while the circles are from the long timescale PDS.
The solid line is the best fit as described in the text, a power-law with
a low-frequency cutoff at $\sim$13~d.
The dotted line is the same power-law without the cutoff.}
\label{fig2}
\end{figure}

\placefigure{fig2}

The combined PDS was then fitted with a model in which a steep power law
dominates
at high frequencies, but cuts off to a slope of $ a = 0 $ at low
frequencies:
\begin{equation}
P(f) = C / ( 1 + f / f_c )^a,
\end{equation}
where $P(f)$ is the fluctuation power at a frequency $f$, $a$ is
the power law slope at high temporal frequencies, $f_c$ is the ``cutoff
frequency," well below which the PDS flattens to a slope of zero, and $C$
is the normalization.
The fit, with three free parameters ($C$, $f_{c}$, and $a$) was a
significant improvement over the simple power law, with $\chi^{2}_{\nu} =
1.70 $ for 12 degrees of freedom, although still formally unacceptable (94\%
confidence) mainly due to the higher absolute level of variability
power in the short time-scale data.
The high-frequency slope of the broken power law fit is $ a = -1.12 $,
similar to the slope derived
using the high-frequency PDS alone, and the cutoff frequency of
8.7~$\times$~10$^{-7}$~Hz corresponds to a timescale of $\sim$13~d.
These parameters are summarized in Table~3.
While the power law plus cutoff model is clearly a marked improvement over the
unbroken power law, we note the exact shape of the turnover is not
well constrained and the low-frequency behaviour not at all
well determined.

\placetable{tab3}

\section{ Discussion and Conclusions }

We have carried out the most comprehensive X-ray variability study of
an NLS1 to
date.
The long and short time-scale light curves of \ark\ are found to be
remarkably similar,
with comparable excess variances over
the sampling intervals of 32 days and 627 days.
The straightforward deduction from a visual examination of the light
curves is that \ark\ exhibits most of its
variability power
on time-scales substantially less than one month. We suggest this 
may well be a common
feature of NLS1s, and one which distinguishes the sub-class from BLS1s.
Turner \et (1999) calculated the excess variance for \asca\
observations of 36 Seyfert galaxies over a typical sampling interval of
1 day finding values up to $\sim$10 times greater for NLS1s compared with
BLS1s (of similar luminosity).
A current explanation for the rapid and large amplitude variability
characteristic of Narrow Line AGN is in terms of a smaller black hole
mass, and hence of size scale. Our result now confirms  - for \ark\ -
that the large short-term variability is indeed primarily
due to the time-scales being shorter (rather than - necessarily
- having greater variability power) than for BLS1 of comparable
luminosity.

The \ark\ PDS now offers an opportunity to
quantify the lower mass and higher accretion rate of a NLS1
in comparison with a BLS1 of similar luminosity. To date the only BLS1
PDS obtained by a similar evenly-sampled campaign is for \ngc, from
which Edelson \& Nandra (1999) obtained a cut-off time-scale
of $\sim$30~d.
Assuming for the moment a simple scaling law applies for the
variability time-scale and black hole
mass of NLS1 and BLS1, our result suggests that
\ark\ has a black hole mass of order $\sim$ 0.4 that of \ngc.
Reverberation mapping by Robinson \et (1994) found \ngc\ to have a
black hole mass of order $ \sim 3 \times 10^7 $~M$_\odot$, implying
a corresponding mass for \ark\ of $ \sim 1 \times 10^7 $~M$_\odot$.
Estimation of the
accretion rate for most AGN is hampered by a large uncertainty
of the luminosity in the hidden EUV band. However, Wandel \et (1999)
showed the bolometric luminosity of BLS1s may be estimated
by scaling from the continuum luminosity at 5100 $\AA$, based on a
comparison of photoionisation and reverberation estimates of
black hole masses. By this means we derive
a bolometric luminosity
for \ngc\ of
$ \sim 10^{44}$~erg/s which should be good to a factor of $ \sim 3$.
The corresponding accretion rate for \ngc\ is then of order 0.05
$\dot{M}_{{\rm Edd}} $.

The very different SED of NLS1s means we cannot rely on the Wandel \et
(1999) relation to estimate the bolometric luminosity of \ark.
Instead we use the respective 2--10keV luminosities, for which
\ark\ is some $\sim$3 times that of \ngc.
Combining the respective estimates of black hole mass and luminosity
then gives an
accretion rate some $\sim$8
times larger for \ark\ than for \ngc, indicating that \ark\ is
accreting at a substantial fraction of the
Eddington limit, or 0.4 $\dot{M}_{{\rm Edd}} $.
We note furthermore that this is
probably a lower limit, given that the scaling with \ngc\ via the 2--10 keV
luminosities takes no account of the strong soft X-ray excess in \ark.

Interpreting the PDS in the above way is subject to two caveats. First,
is the variability stationary, to at least the extent that
the measured ''characteristic time-scale'' is a reliable measure
of scale and black
hole mass? The uniformity of the long time-scale light curve of \ark\
shown in Figure 1
is reassuring in that respect, suggesting the absence of low frequency
power is sustained throughout our campaign. On the other hand, the
higher mean flux and larger fractional variability during
the intensive monitoring suggests a degree of non-stationarity, but
not to an extent which alters our basic finding, 
that \ark\ contains
most of its variability power on timescales much shorter than 1 month.
A second question is whether the accretion discs in both \ark\ and \ngc\
extend close to the innermost stable orbit, justifying our inverse 
scaling of black hole mass with
the respective values of $ f_c $?
The high luminosity of NLS1s indicate that is generally true
for NLS1s, while
in the case of \ngc\
the observation of a broad Fe-K fluorescence line (Nandra \et 1999)
suggests that its accretion disc also extends
close to the
innermost stable orbit.
 
We conclude that comparisons of the X-ray
variability of \ark\ and \ngc\
indicate a black hole mass for \ark\ of $ \sim 1 \times
10^7 $~M$_\odot$, implying an accretion rate in the range 0.2--1.0
$\dot{M}_{{\rm Edd}} $. It is interesting to compare this result with
the accretion rate derived from modelling of line widths and
variability time-scales in the optical Broad Line
Region (Laor 1998). Taking the above estimate for the luminosity of \ark\
and H$\beta$ FWHM $ = 800 $~km/s (Vaughan \et 1999b), the relation
in Laor (1998)
yields an ''independent" estimate for \ark\ of 1.0
$\dot{M}_{{\rm Edd}} $

\acknowledgements
The authors acknowledge the dedicated work of the \xte\ team, especially Evan
Smith for his careful scheduling so close to the ephemeris.
RAE and AM were supported by NASA grants NAG 5-7315 and NAG
5-9023, and SAV by
the UK Particle Physics and Astronomy Research Council.

\pagebreak

\begin{deluxetable}{ccccccc}
\tablewidth{0pc}
\tablenum{1}
\tablecaption{Sampling Parameters \label{tab1}}
\small
\tablehead{
\colhead{Time} &
\colhead{Observing Dates} &
\colhead{Mean Sampling} &
\colhead{Data} &
\colhead{No. of Flux} & \colhead{Usable Temporal} \\
\colhead{Scale} & \colhead{(MJD)} &
\colhead{Interval} &
\colhead{Lost} &
\colhead{Points} & \colhead{Frequency Range} }
\startdata
Long   &  51179.55--51806.78 & 4.267~d & 9 points  & 148 &
6.9~$\times$~$10^{-8}$~--~1.1~$\times$~$10^{-6}$\\
Short  &  51694.82--51726.47 & 3.2~hr  & 13 points & 235 &
1.4~$\times$~$10^{-6}$~--~3.4~$\times$~$10^{-5}$\\
\enddata
\end{deluxetable}

\begin{deluxetable}{cccccc}
\tablewidth{0pc}
\tablenum{2}
\tablecaption{Multi-Band Light Curve Properties \label{tab2}}
\small
\tablehead{
\colhead{Time} &
\colhead{Mean} &
\colhead{Excess} &
\colhead{Power-Law} \\
\colhead{Scale} &
\colhead{ct~s$^{-1}$} &
\colhead{Variance} &
\colhead{Slope} }
\startdata
Long    & $1.63~\pm~0.04$  & $0.076~\pm~0.009$ & $-0.24~\pm~0.08$ \\
Short   & $1.87~\pm~0.04$  & $0.111~\pm~0.010$ & $-0.96~\pm~0.07$  \\
\enddata
\end{deluxetable}

\begin{deluxetable}{ccccc}
\tablewidth{0pc}
\tablenum{3}
\tablecaption{PDS Fit Parameters \label{tab3}}
\tablehead{
\colhead{Description} & \colhead{Parameter} & \colhead{value}}
\startdata
High-Frequency Slope & $a$   &  $-1.12~\pm~0.04$\\
Cutoff Frequency & $f_{c}$ &  $8.72~\pm 0.81 \times 10^{-7}$~Hz\\
Cutoff Timescale & $t_{c}(=1/f_{c})$  & $13.3^{+1.3}_{-1.2}$ days\\
Normalization Coefficient & $C_{1}$ &  $4.4\pm0.2 \times 10^{6}$~Hz$^{-1}$\\
\enddata
\end{deluxetable}

\pagebreak

\pagebreak

\begin{references}

\reference{B96} Boller,Th.,Brandt,W.N.\ \& Fink,H.\ 1996, A\&A,
305, 53

\reference{B97} Brandt,W.N.,Mathur,S.\ \& Elvis,M.\ 1997, MNRAS, 285,
25P

\reference{B99} Brandt, W.N.,Boller,Th.,Fabian,A.C.\ \& Ruszkowski,
M.\ 1999, MNRAS,303,L53

\reference{B81} Brillinger, D., 1981 ``Time Series: Data Analysis and
Theory,'' 2nd Edition (Holden-Day Publishing)

\reference{E96} Edelson, R.\ \& Nandra, K.\ 1999, \apj, 514, 682

\reference{G89} Goodrich, R.W.\ 1989, \apj, 342, 234

\reference{J96} Jahoda, K., Swank, J., Giles, A., Stark, M.,
        Strohmayer, T., Zhang, W., Morgan, E.H.\ 1996, SPIE 2808, 59

\reference{L97} Laor, A. \et \ 1997, \apj, 477, 93

\reference{L98} Laor, A.\ 1998, \apj, 505, L83

\reference{L99} Leighly, K.\ 1999, ApJS, 125, 297

\reference{L93} Lawrence, A.\ \& Papadakis, I.\ 1993, \apj, 414, L85

\reference{M01} Markowitz, A.\ \& Edelson, R.\ 2001, \apj, in press

\reference{M88} McHardy, I.\ 1988, Mem Soc Ast Ital, 59, 239

\reference{N97} Nandra, K.,George,I.M.,Mushotzky, R.F.,Turner,T.J.\ \&
Yaqoob, T.\ 1997, \apj, 476, 70

\reference{N99} Nandra, K.,George,I.M.,Mushotzky, R.F.,Turner,T.J.\ \&
Yaqoob, T.\ 1999, \apj, 523, L17

\reference{O75} Oppenheim, A.\ \& Shafer, R.\ 1975, ``Digital Signal
Processing,'' (Prentice-Hall Publishing)

\reference{O85} Osterbrock, D.E.\ \& Pogge, R.W.\ 1985, \apj, 297, 166

\reference{PM95} Papadakis, I.\ \& McHardy, I.\ 1995, MNRAS, 273, 923

\reference{P95} Pounds,K.A., Done,C.\ \& Osborne, J.\ 1995, MNRAS,
277, L5

\reference{R94} Robinson, A.\ 1994, ASP Conf Ser, 69, 147

\reference{T99} Turner, T.J.,George,I.M.,Nandra,K.\ \& Turcan,D.\ 1999,
\apj, 524, 667

\reference{V99a} Vaughan, S.,Pounds,K.A.,Reeves,J.N.,Warwick,R.S.\ \&
Edelson,R.A.\ 1999a, MNRAS, 308, L34

\reference{V99b} Vaughan, S.,Reeves,J.N.,Warwick,R.S.\ \&
Edelson,R.A.\ 1999b, MNRAS, 309, 113

\reference{W99} Wandel, A.,Peterson,B.M.\ \&
Malkan,M.A.\ 1999, MNRAS, \apj, 526, 579

\end{references}
\end{document}